\documentclass[conference]{IEEEtran}
\IEEEoverridecommandlockouts
\usepackage{cite}
\usepackage{amsmath,amssymb,amsfonts}
\usepackage{algorithmic}
\usepackage{graphicx}
\usepackage{textcomp}
\usepackage{xcolor}
\usepackage{bm}
\usepackage{multirow}
\usepackage{wrapfig,lipsum,booktabs}
\usepackage{hyperref}
\def\BibTeX{{\rm B\kern-.05em{\sc i\kern-.025em b}\kern-.08em
    T\kern-.1667em\lower.7ex\hbox{E}\kern-.125emX}}
\begin{document}

\title{Influences of Temporal Factors on GPS-based Human Mobility Lifestyle\\
}

\author{\IEEEauthorblockN{Tran Phuong Thao}
\IEEEauthorblockA{\textit{The University of Tokyo} \\
Tokyo, Japan \\
 tpthao@yamagula.ic.i.u-tokyo.ac.jp}
}

\maketitle

\begin{abstract}
Analysis of human mobility from GPS trajectories becomes crucial in many aspects such as policy planning for urban citizens, location-based service recommendation/prediction, and especially mitigating the spread of biological and mobile viruses. In this paper, we propose a method to find temporal factors affecting the human mobility lifestyle. We collected GPS data from 100 smartphone users in Japan. We designed a model that consists of 13 temporal patterns. We then applied a multiple linear regression and found that people tend to keep their mobility habits on Thursday and the days in the second week of a month but tend to lose their habits on Friday. We also explained some reasons behind these findings.

\end{abstract}


\section{Introduction}
Understanding individual human mobility plays an important role especially when the geographic spread of the infectious virus that causes COVID-19 has taken the world into uncharted territory. Not only that, it is also a critical factor in policy planning~\cite{Geography2001, Cities2015}, travel demand forecasting~\cite{Transportation2000, Tourism2017}, location-based recommendation/service advertising~\cite{KDD12}, or location-based personal authentication~\cite{thaodbsec}. M. Gonzalez et al.~\cite{MIT2008} proved that human mobility follows a high degree of regularity. Therefore, several sophisticated models have been proposed to determine the factors influencing the probability that people tend to increase and lose their mobility lifestyle. The factors can be classified into spatial, temporal, and social in which temporal one has been proved to be the most important affecting factor. However, the temporal factors found in existing work are still coarse-grained (i.e., weekend/weekday without clarifying which specific days of the week, or which week of the month, etc.)

In this paper, we investigated the recurrence and temporal periodicity inherent to
human mobility inferred from mobile phone data with more fine-grained factors. We collected GPS data from 100 random smartphone users in Japan. We designed a model consisting of 13 temporal factors from 3 pattern categories (i.e., days of the weeks, quarters of the month, and holidays including weekend and national public holidays) for independent variables. We also proposed an algorithm to compute the probability (i.e., similarity score) of the users to visit the locations they visited before for the target outcome. We then applied a multiple linear regression and performed a $t$-test. We found that people tend to keep their mobility habits on Thursday and the days in the second week of a month but tend to lose their habits on Friday. We also discussed some reasons and applications behind these findings. 

The rest of this paper is organized as follows. Section~\ref{section:relatedwork} introduces related work. Section~\ref{section:method} presents our proposed methodology. Section~\ref{section:experiment} gives the experiment and our findings. Section~\ref{section:discussion} discusses applications and limitations of our method. Section~\ref{section:conclusion} describes the conclusion.

\section{Related Work}
\label{section:relatedwork}
In this section, we introduce related work about factors affecting the location habit. The work can be classified into three research directions.

\subsection{Spatial Factors}
S. Zhao et al.~\cite{AAAI16} observed that 80\% successive checked-in POIs (Points-of-Interest) happen in less than 32 kilometers. They explained that people often act around their home or office, so even being independent with the last check-in, the successive check-in can still happen in the same activity area. S. Yali et al.~\cite{KBS2019} analyzed the two location-based social networks Foursquare and Gowalla. They found that the probabilities for distances within 5 km are greater than 40\%, which decrease to17\% and 8\% within 10 km on the datasets, respectively. Most users checked in within 20 km. T. Thao et al.~\cite{DPMarxiv20, DPM20} leveraged the idea that the locations at close time clocks have a closer correlation in physical distance than the locations at far time clocks since a human needs a period of time to move from a location to another location gradually. The experimental result showed that the extracted distance coherence features along with the longitudes and latitudes could improve the authentication's accuracy. While all the papers~\cite{AAAI16, KBS2019, DPM20} focused on the fact that closer locations have a higher probability of being visited by users, Y. Hongzhi et al.~\cite{KDD13} raised a more challenging problem when people travel to a new city where they have no activity history. They showed that people tend to travel a limited distance when visiting venues and attending events. Furthermore, the activity records in their non-home cities are only 0.47\% of the activity records when living in their home cities. To solve the problem, the authors analyzed the two factors including user interest (e.g., kids would pay more attention in playgrounds while young ladies may be more interested in cosmetics stores) and local preference (e.g., people are more likely to visit local sightseeing attractions and attend popular events in the city when they travel to an unfamiliar city). They found that the factors also affect the decision to visit an unfamiliar location.

\subsection{Temporal Factors}
G. Huiji et al.~\cite{temporal2013} extracted the correlations between the check-in time and the corresponding check-in preferences of a user. They found that weekly patterns (7 days of the week) and weekday/weekend patterns can capture the temporal check-in preferences of a user. However, the results do not clearly indicate which day of the week, weekday, or weekend is the affecting factor but only the general patterns. S. Zhao et al.~\cite{WWW17} found that the day of week check-in pattern at different hours: users take more check-ins in the late afternoon and the evening from 04:00 p.m. to 3:00 a.m. on weekends than the weekdays. Saturday and Sunday take a similar pattern, while the days from Monday to Friday take a similar pattern that is different from the weekends. It may infer that weekday and weekend are two types of effects on the check-in behavior of the user. J. Bao et al.~\cite{SIGSPATIALSpecial2016} split a week into two parts including weekdays and weekends. For each part, they split a day into hourly time bins. A total of $24\times2$ time bins are used for the expression of temporal patterns. M. Gonzalez et al.~\cite{MIT2008} measured the return probability for each individual. They found that the probability that a user returns to the position where the user was first observed after $t$ hours for a two-dimensional random walk should follow $\frac{1}{t \ln^{2}(t)}$. The return probability is characterized by several peaks at 24h, 48h, and 72h, which indicates a strong tendency of humans to return to locations they visited before. M. Xie et al.~\cite{CIKM16} explored the importance of spatial, temporal, and social factors and found that they can be ranked as follows: temporal effect $>$ content effect $>$ spatial effect. This indicates the temporal factors may provide the most information although of course, combining them is the best solution. 

\subsection{Social and Content Factors}
H. Wang et al.~\cite{SIGSPATIAL13} studied that the social link is an important factor affecting the choices of people when deciding which new place to visit. The authors analyzed the Gowalla dataset and found that a friend or a friend-of-a-friend has visited more than 30\% of the new places visited by a user in the past. With the same observation that social friends tend to have similar check-in behavior, several papers~\cite{SIGSPATIALZhang13, VLDB2017, MobiQuitous09, ASONAM15} also extracted the similarity score between the users derived from the social friendships. The experiment result showed that it could enhance the accuracy. Besides the links of friend and friend-of-a-friend, H. Bagci et al.~\cite{WWW16} showed that local expert is also a factor affecting the place to visit. J. Bao et al.~\cite{expert12} pointed out that users who visit many high-quality locations tend to have high knowledge about the vicinity. In a similar manner, if a particular location is visited by many high-quality users (i.e., experts), it is more probable for that location to be a quality location. L. Kai et al.~\cite{advertise2012} aimed at the service locations only, such as restaurants, fitness centers, etc. They found that the factors including demographics, preferences, and service levels (e.g., price range, discount or not, advertisements) can increase the probability of mobile users visiting the service locations.

\section{Methodology}
\label{section:method}
In this section, we present our proposed methodology including data collection and the model design.

\subsection{Data Collection}
A navigation application named MITHRA (Multi-factor Identification/auTHentication ReseArch) was created to collect the GPS information from the Android smartphone users.  
One hundred users were randomly recruited, thus live and work in random areas. The data consists of timestamps and GPS information (longitude and latitude). The application collects the data every 5 minutes. The users have different data collection periods because it depends on the time that each user stars running the application. The entire collected data from all the users is from January to April 2017. The timestamp is up to seconds. The precision of the longitudes and latitudes is six decimal places (e.g., 36.xxxxxx), which correspond to 0.1 meters. Regarding the data privacy, a privacy consent is shown to the users during the installation process. The application can only be successfully installed if the users accept the terms and conditions agreement. We do not collect any personal information such as name, age, date of birth, gender, etc. except address which is used for user identity. Our project is reviewed by the Ethics Review Committee of the Graduate School of Information Science and Technology, the University of Tokyo. 

\begin{table*}[!ht]
\centering
\caption{Example of Similar Score Calculation for An User $U$}
\begin{tabular}{cc|cc|cc|c|cc}
\multicolumn{2}{c}{\textbf{H}}  & \multicolumn{2}{c|}{\textbf{00:00-00:59}} & \multicolumn{2}{c|}{\textbf{01:00-01:59}} & $\cdots$ &  \multicolumn{2}{c}{\textbf{23:00-23:59}} \\
\hline
\multirow{4}{*}{$D_{learn}$} & & $U_{learn\_0}$ & $W_{learn\_0}$ &  $U_{learn\_1}$ & $W_{learn\_1}$ & $\cdots$ &  $U_{learn\_{23}}$ & $W_{learn\_{23}}$ \\
\cline{3-9}
& & ($lon_{1}, lat_{1}$) & \color{red} $\bm{weight_{1}}$ & ($lon_{3}, lat_{3}$) & \color{red} $\bm{weight_{3}}$ & $\cdots$ & ($lon_{30}, lat_{30}$) & \color{red} $\bm{weight_{30}}$ \\
& & ($lon_{2}, lat_{2}$) & \color{red} $\bm{weight_{2}}$ & ($lon_{4}, lat_{4}$) & \color{red} $\bm{weight_{4}}$ & $\cdots$ & ($lon_{31}, lat_{31}$) & \color{red} $\bm{weight_{31}}$ \\
& &  &  & ($lon_{5}, lat_{5}$) & \color{red} $\bm{weight_{5}}$ & $\cdots$ & &  \\
\hline
\multirow{4}{*}{$D_{test}$} & date & $r_{1}$ & $s_{1}$ & $r_{2}$ & $s_{2}$ & $\cdots$ & $r_{23}$ & $s_{23}$ \\
\cline{2-9}
& 2017/04/01 & ($lon_{1}, lat_{1}$) & \color{red} $\bm{weight_{1}}$ & ($lon_{6}, lat_{6}$) & \color{red} \textbf{0} & $\cdots$ & ($lon_{31}, lat_{31}$) & \color{red} $\bm{weight_{31}}$\\
& 2017/04/02 & ($lon_{10}, lat_{10}$) & \color{red} \textbf{0} & ($lon_{5}, lat_{5}$) & \color{red} $\bm{weight_{5}}$ & $\cdots$ & ($lon_{31}, lat_{31}$) & \color{red} $\bm{weight_{31}}$\\
& 2017/04/03 & ($lon_{10}, lat_{10}$) & \color{red} \textbf{0} & ($lon_{3}, lat_{3}$) & \color{red} $\bm{weight_{3}}$ & $\cdots$ & ($lon_{32}, lat_{32}$) & \color{red} \textbf{0}
\end{tabular}
\label{table:example}
\end{table*}

\subsection{Model}
At first, we briefly describe how a linear regression work. Linear regression is a statistical method used for measuring whether a set of factors affect (or can be used to predict) a certain outcome. It can model the relationship between one or more independent variables (features) and one dependent (output) variable. The value of the target function is expected to be a linear combination of the features. Formally, let $f^{*}$ denote the predicted value:
\begin{equation}
f^{*}(c, x) \sim c_{0} + c_{1}x_{1} + \cdots +  c_{n}x_{n}	
\end{equation}
where $X = \{x_{1}, x_{2}, \cdots, x_{n}\}$ denotes the set of features, $n$ denotes the number of features, $C = \{c_{1}, c_{2}, \cdots, c_{n}\}$ denotes the set of coefficients, and $c_{0}$ denotes the intercept. $c_{0}$ is a constant representing the expected mean value of $f^{*}$ when $x_{i} = 0$ for all $i = \{1, \cdots, n\}$. There are several methods to solve the regression (e.g., Ridge Regression, Lasso, etc.) but we use the most common method Ordinary Least Squares (OLS) which minimizes the residual sum of squares between the observed targets in the dataset, and the targets predicted by the linear approximation:
\begin{equation}
	min_C ||Xc-f||_2^2	
\end{equation}
When $x_{1}, x_{2}, \cdots, x_{n}$ are correlated and the columns of the design matrix $X$ are approximately linear dependent, $X$ will become close to singular. 

We are now ready to define our model for the regression. For each user $U$, the model is defined as:
\begin{equation}
score \sim  wdays + mquar + hdays 
\end{equation}
where $score$ represents the target function; $wdays$,  $mquar$, and $hdays$ represent the variables related to the days of the weeks, quarters of the month, and holidays, respectively. 

\subsubsection{Target Function (Dependent Variable)}
In this part, we explain the algorithm used to calculate the similarity score, which measures the probability of a user re-visiting a location that he/she visited before. The scores also represent the mobility lifestyle pattern of a user. For each user $U$, the data is splitted into two parts based on the data collection time period. The data from the first half of the time period is denoted by $D_{learn}$ and the one from the later half is denoted by $D_{test}$. The similarity score between $D_{learn}$ and $D_{test}$ is used for the target function. The procedure to calculate the similarity score is described as follows. 

\paragraph{Measuring Template from $D_{learn}$}
First, the longitude and latitude in each data record $d_{i} \in D_{learn}$ are rounded to 2 decimal places from original 6 decimal places since the location accuracy of people's movement is often within 1 km square. Let $dat_{i}, tim_{i}, lon_{i}, lat_{i}$ denote the date (year, month, day), the time (hour, minute, second), the longitude and latitude after being rounded, of $d_{i}$, respectively. Let $H = \{\text{00:00-00:59}, \text{01:00-01:59}, \cdots, \text{23:00-23:59}\}$ be the 24 hourly-time periods. Each period is denoted by $h_{\alpha} \in H$ where $\alpha \in [0, 23]$. The records in $D_{learn}$ are grouped into 24 subsets according to $h_{\alpha}$. For each $\alpha$, the following sets are constructed:
\begin{itemize}
\item $T_{learn\_\alpha} = \{(lon_{i}, lat_{i})\}$: the set contains the longitude and latitude of all the records $d_{i}$ such that $tim_{i} \in h_{\alpha}$ regardless of $dat_{i}$.
\item $U_{learn\_\alpha} = \{(lon\_uniq_{j}, lat\_uniq_{j})\} \subset T_{learn\_\alpha}$: the set contains only the unique pairs of longitude and latitude. For $\forall j, j' \in [0, \mid U_{learn\_\alpha}\mid]$, $(lon\_uniq_{j} \ne lon\_uniq_{j'}) \lor (lat\_uniq_{j} \ne lat\_uniq_{j'})$ (remark, it is an OR, not AND operation).
\item $W_{learn\_\alpha} = \{weight_{j}\}$: the set contains the corresponding weight of the pair $(lon\_uniq_{j}, lat\_uniq_{j}) \in U_{learn\_\alpha}$. $U_{learn\_\alpha}$ and $M_{learn\_\alpha}$ have the same length. The weight is calculated as the percentage that the user $U$ stays at the coordinate $(lon\_uniq_{j}, lat\_uniq_{j})$, that is the ratio between the number of the pair values $(lon\_uniq_{j}, lat\_uniq_{j})$ and the length of $T_{learn\_\alpha}$:
\begin{equation}
weight_{j} = \frac{\#(lon\_uniq_{j}, lat\_uniq_{j})}{\mid T_{learn\_\alpha}\mid}
\end{equation}
\end{itemize}

\paragraph{Extracting Representatives from $D_{test}$}
In $D_{learn}$, we grouped the data into 24 hours regardless of the date. For $D_{test}$, we consider each different date before grouping the data of the date into 24 hours. For each unique date $\delta$ from the data in $D_{test}$ and for each $\alpha \in [0, 23]$, we also construct $T_{test\_\alpha} =  \{(lon_{i}, lat_{i})\} $ in the same way as $T_{learn\_\alpha}$ but with $dat_{i} = \delta$. We determine the representative $r_{test\_\delta\alpha}$ for $T_{test\_\alpha}$ by extracting the element $(lon_{i}, lat_{i}) \in T_{test\_\alpha}$ at which the user $U$ stays for the longest period of time on the date $\delta$. We have $\alpha$ representatives for entire $D_{test}$. 

\paragraph{Matching to Calculate Similarity Scores}
For each date $\delta$ in $D_{test}$ and for each $\alpha \in [0, 23]$, if the representative $r_{\delta\alpha}$ exists in $U_{learn\_\alpha}$, the similarity score $s_{\delta\alpha}$ will be set to the corresponding weight from $W_{learn\_\alpha}$. If not, $s_{\delta\alpha}$ is set to zero. The example is given in Table~\ref{table:example}. After all the weights for 24 hours in each day $\delta$ are computed, all the weights in $D_{test}$ for the user $U$ are summed up and used for the final value of $score$. So, each user $U$ has a corresponding similarity score. For the example in Table~\ref{table:example}, the final score for $U$ is $weight_{1} + weight_{5} + weight_{3} + \cdots + 2weight_{31}$.

\subsubsection{Variables}
\label{section:variables}
For each user $U$ and each day $\delta$ mentioned above, the following binary variables were extracted. The first group is 7 binary variables which correspond to 7 days of the week (i.e.,  is $\delta$ Monday, $\cdots$, is $\delta$ Sunday) denoted by $\{\mathsf{mon}$, $\mathsf{tue}$, $\cdots$, $\mathsf{sun}\}$. The second group is 4 variables which correspond to 4 weeks of the month (i.e., is $\delta$ the first week, $\cdots$, is $\delta$ the fourth week) denoted by $\{\mathsf{wk1}, \mathsf{wk2}, \mathsf{wk3}, \mathsf{wk4}\}$. The third group is 2 variables related to holidays (i.e., is $\delta$ a weekend and is $\delta$ a national holiday) denoted by $\{\mathsf{natl}, \mathsf{wknd}\}$. These 13 binary variables are summed up for all the days $\delta$ of each user $U$. $wdays$,  $mquar$, and $hdays$ represent the summed variables for the first, second, and third group, respectively. Let $D_{P}$ denote the final data which will be used for the regression which consists of 100 samples with 13 variables. 

\begin{table}[!ht]
\centering
\caption{Variables Distribution}
\begin{tabular}{c l rrrrrr}
\textbf{no} & \textbf{var.} & \textbf{mean} & \textbf{SD} & \textbf{kurtosis} & \textbf{skew} & \textbf{min} & \textbf{max}\\
\hline
1 & $\mathsf{mon}$ & 4.42 & 0.57 & -0.76 & -0.36 & 3 & 5\\
2 & $\mathsf{tue}$ & 4.16 &  0.42 & 1.28 & 1.01 & 3 & 5\\
3 & $\mathsf{wed}$ & 4.16 & 0.42 & 1.28 & 1.01 & 3 & 5\\
4 & $\mathsf{thu}$ & 4.18 & 0.46 & 0.66 & 0.67 & 3 & 5\\
5 & $\mathsf{fri}$ & 4.25 & 0.46 & -0.42 & 0.85 & 3 & 5\\
6 & $\mathsf{sat}$ & 4.28 & 0.62 & 0.75 & -0.53 & 2 & 5\\
7 & $\mathsf{sun}$ & 4.57 & 0.50 & -1.96 & -0.29 & 4 & 5\\
8 & $\mathsf{wk1}$ & 6.69 & 1.65 & 8.00 & -2.05 & 0 &11\\
9 & $\mathsf{wk2}$ & 8.00 & 1.21 & 9.12 & 2.42 & 5 & 13\\
10 & $\mathsf{wk3}$ & 8.05 & 1.40 & 8.87 & 1.73 & 4 & 15\\
11 & $\mathsf{wk4}$ & 8.88 & 1.66 & 2.78 & -1.32 & 2 & 12\\
12 & $\mathsf{natl}$ & 0.89 & 0.31 & 4.5 & -2.53 & 0 & 1\\
13 & $\mathsf{wknd}$ & 8.85 & 0.82 & -0.67 & -0.16 & 7 & 10\\
14 & $\mathsf{score}$ & 459.38 & 140.25 & -0.42 & -0.44 & 102.63 & 708.04\\
\end{tabular}\\ $ $\\
Abbreviation: SD (standard deviation), var. (variables)
\label{table:distribution}
\end{table}

\begin{table}[!ht]
\centering
\caption{Jarque-Bera Test for Residuals}
\begin{tabular}{l | r r r r}
\textbf{metrics} & \textbf{entire} & \textbf{excluded} & \textbf{excluded} \\
 &  \textbf{100 Samples} & \textbf{(1, 21, 94)} & \textbf{(1, 21, 30)} \\
\hline
\#samples & 100 & 97 & 97\\
Jarque-Bera score& 7.27 & 5.97 & 6.05 \\
$p$-value & 0.03 & 0.05 & 0.05\\
kurtosis & -0.64 & -0.57 & -0.59\\
skew & 2.64 & 2.57 & 2.69\\
\end{tabular}
\label{table:residuals}
\end{table}

\section{Experiment}
\label{section:experiment}
The program is written in Python 3.7.4 on a computer MacBook Pro 2.8 GHz Intel Core i7, RAM 16 GB. The multiple (linear) regression model is executed using \emph{scikit-learn} package version 0.21. The t-test is computed using \emph{statsmodels} package version 0.11.

\subsection{Distribution of Variables and Normality of Residuals}
The distribution of the 13 variables and the target score is given in Table~\ref{table:distribution}. While the independent variables ($wdays$, $mquar$, $hdays$) and dependent variable ($score$) do not need to be normally distributed, the normality is required for the residuals. The entire preprocessed data ($D_{P}$ as mentioned in Section~\ref{section:variables}) has 100 samples corresponding to 100 users with 13 variables. We performed an \emph{Jarque-Bera test}, and the result is showed in the second column in Table~\ref{table:residuals}. The $p$-value is less than 0.05, which indicates that the residuals are not normally distributed. Therefore, we had to conduct an analysis of data outliers in the next part. 

\begin{figure}[!ht]
\centering
\includegraphics[scale=0.38]{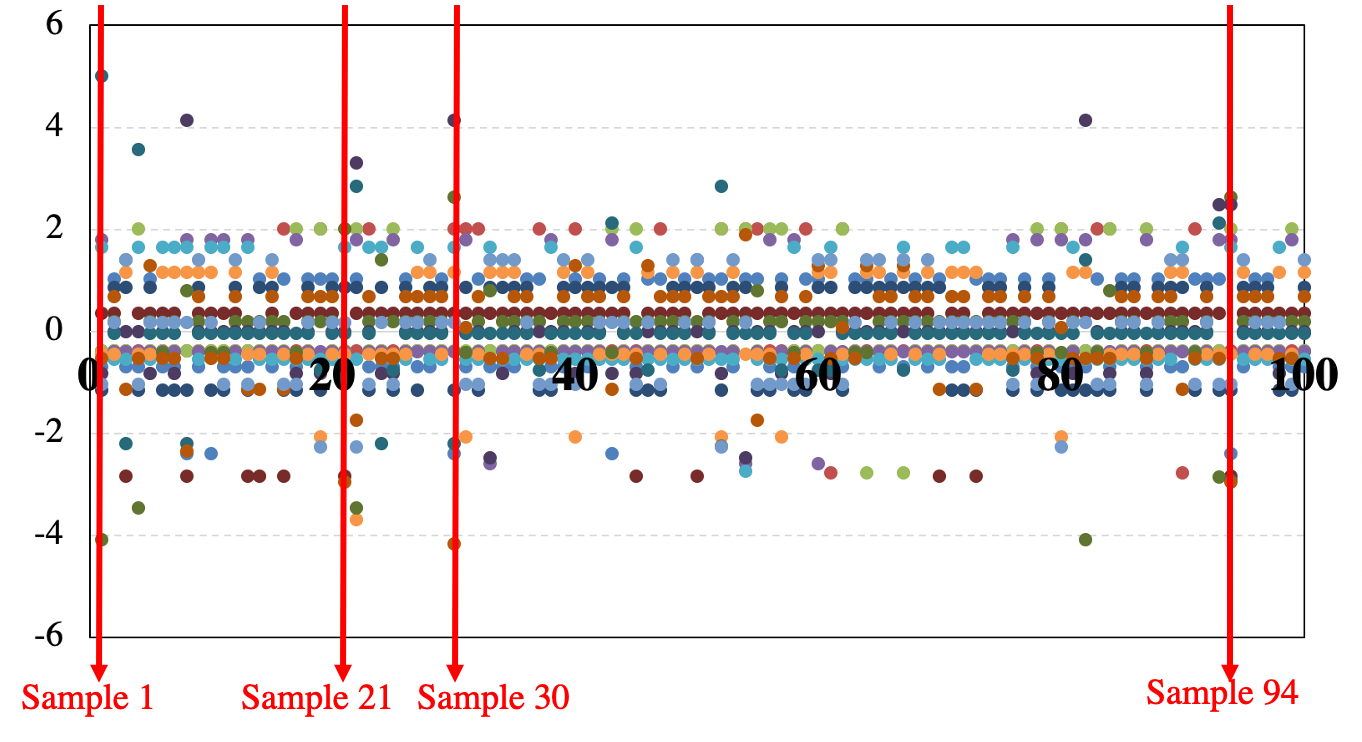}
\caption{Z-Sscore Plotting for 100 Datapoints}
\label{fig:zscore}
\end{figure}

\begin{figure*}[!htb]
\minipage{0.333\textwidth}
  \includegraphics[scale=0.38]{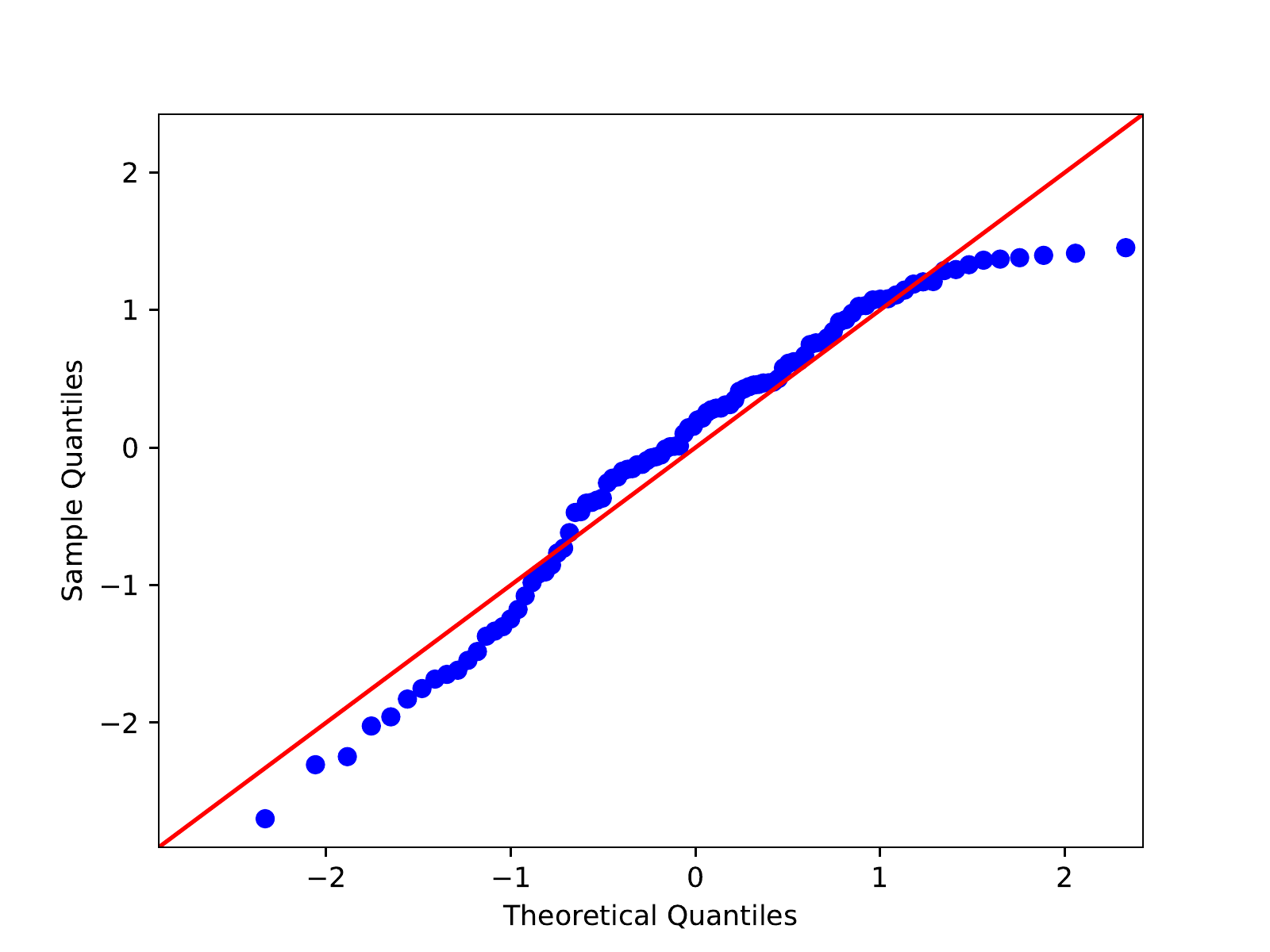}
  \caption{QQ-Plot of Entire Data}
  \label{fig:qqlot_entire}
\endminipage\hfill
\minipage{0.333\textwidth}
  \includegraphics[scale=0.38]{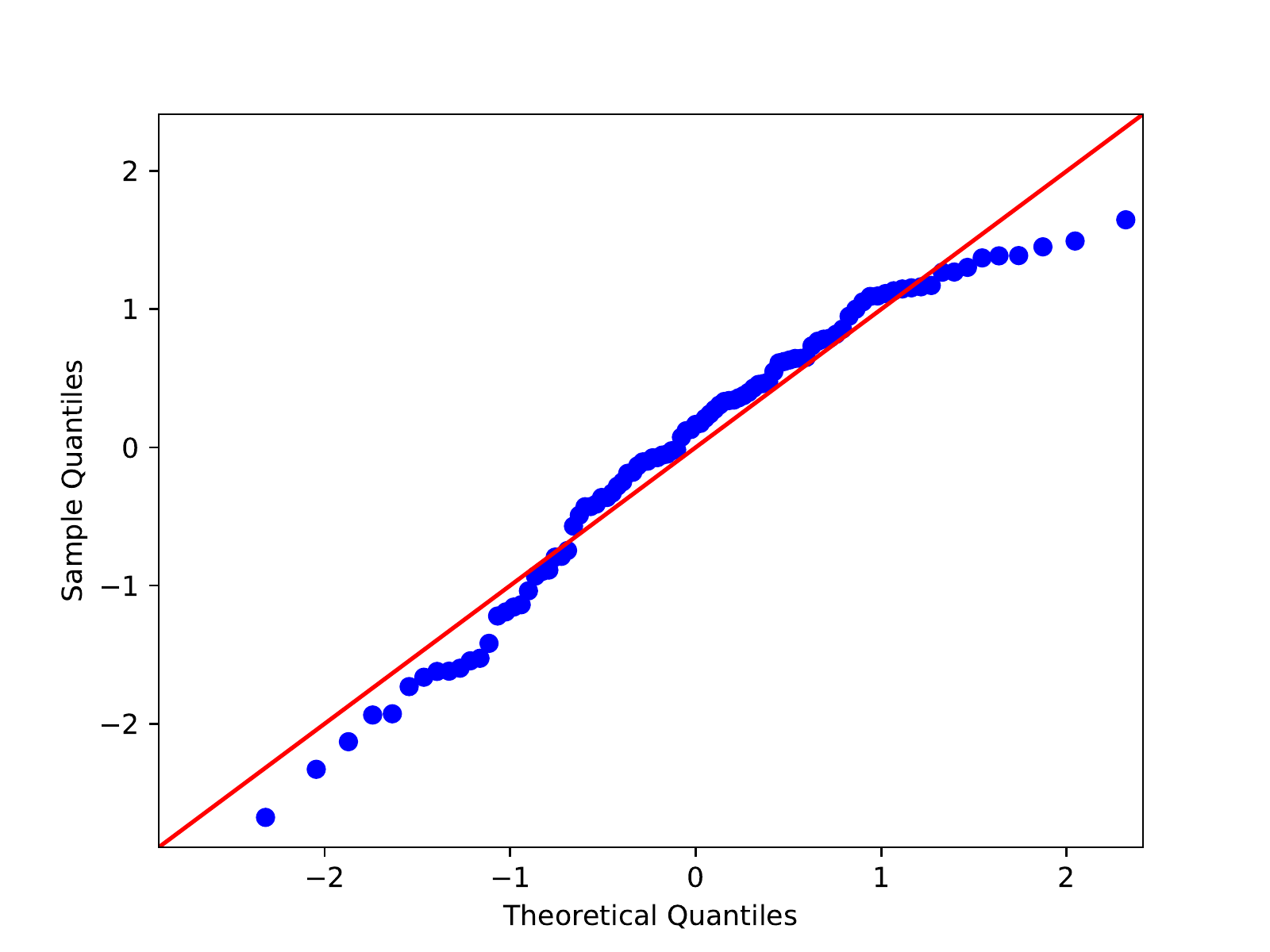}
  \caption{QQ-Plot After Removing (1, 21, 94)}
  \label{fig:qqlot_case1}
\endminipage\hfill
\minipage{0.333\textwidth}%
  \includegraphics[scale=0.38]{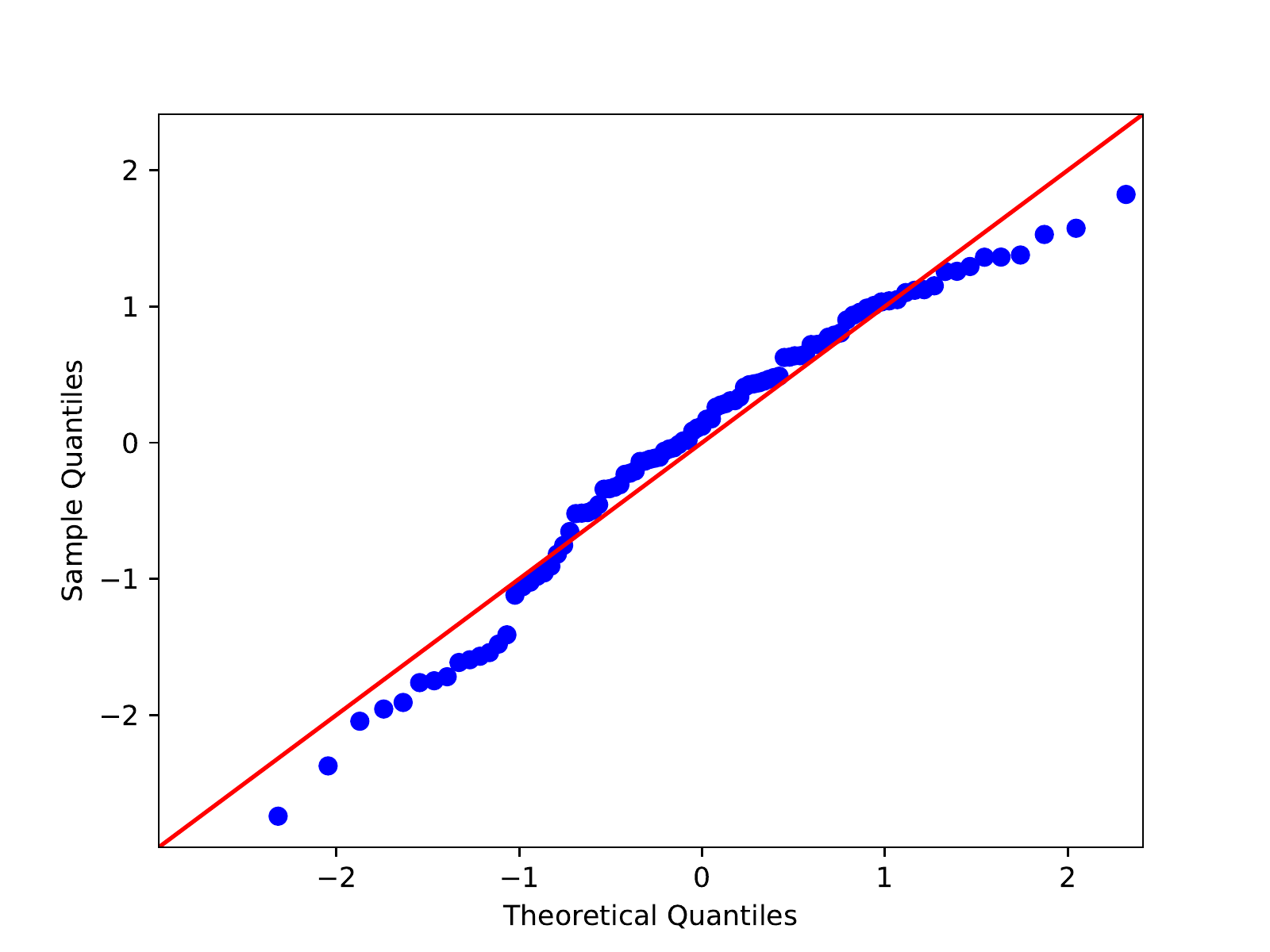}
  \caption{QQ-Plot After Removing (1, 21, 30)}
  \label{fig:qqlot_case2}
\endminipage
\end{figure*}

\subsection{Outlier Identification}
First, we measured the $z$-score for each of the 13 variables from 100 samples. According to the \emph{empirical rule} (so-called \emph{68--95--99.7 rule} or \emph{three-sigma rule})~\cite{EmpiricalRule2006}, any $z$-score that is greater than 3 or less than -3 is considered to be an outlier. Almost all of the data (99.7\%) should be within three standard deviations from the mean; and 99.7\% of the z-scores to be within the range (-3, +3).  Therefore, we scanned all the $z$-scores and could find six samples that have any of 13 variables with $z$-score greater than 3 or less than -3. The 6 outliers are the 1st, 4th, 8th, 22nd, 30th, and 82nd sample in $D_{P}$ denoted by $outlier_{(-3, +3)} = \{s_{1}, s_{4}, s_{8}, s_{22}, s_{30}, s_{82}\}$. Our aim is to remove the smallest number of outliers such that the $p$-value of the residuals can be increased up to 0.05 or more. We, therefore, run an algorithm to perform the Jarque-Bera test after removing each $k$-combination of the elements in the set $outlier_{(-3, +3)}$. $k$ is chosen in ascending order from 1 to the length $n = |outlier_{(-3, +3)}| = 6$. Remark that we do not need to check all the combinations $\sum^{n}_{k=1} {n \choose k}$. If we can find a $p$-value that is equal or greater than 0.05 at a certain $k = k_{p}$, it is unnecessary to check the other combinations with $k>k_{p}$. Unfortunately, we could not find (to remove) any outlier combination that can pass the Jarque-Bera test. 

Therefore, we then reduced the outlier range from (-3, +3) to (-2.9, +2.9) and could extract 8 samples, says $outlier_{(-2.9, +2.9)} = \{s_{1}, s_{4}, s_{8}, s_{21}, s_{22}, s_{30}, s_{82}, s_{94}\}$. Similarly, we also performed the Jarque-Bera test; and fortunately, we could find two combinations at $k = 3$ that can boost the $p$-value when removing them: $C_{1} = \{1, 21, 94\}$ and $C_{2} = \{1, 21, 30\}$. The $z$-scores of all 100 samples are plotted in Fig.~\ref{fig:zscore}. 13 colors of the datapoints represent 13 variables. All the data belonging to the 4 outlier samples $s_{1}$, $s_{21}$, $s_{30}$, and $s_{94}$ lie along the 4 red lines. The results of the tests are summarized in the last two columns of Table~\ref{table:residuals}. Let $D_{C1} = D_{P} \setminus C_{1}$ and $D_{C2} = D_{P} \setminus C_{2}$ denote the data after removing the outliers from $C_{1}$ and $C_{1}$.  The Quantile-Quantile (QQ) plots of $D_{P}$, $D_{C1}$, and $D_{C1}$ are given in Figures~\ref{fig:qqlot_entire}, ~\ref{fig:qqlot_case1}, and~\ref{fig:qqlot_case2}, respectively. It can observe that the datapoints from $D_{C1}$ and $D_{C1}$ are closer to the straight 45-degree reference lines than those from $D_{P}$. 

It may raise the question of why not just remove all the data outliers. First, we should note that removing all the outliers does not mean that the $p$-value for the residuals can be increased. We made a test when removing the 6 samples from $outlier_{(-3, +3)}$ and the 8 samples from $outlier_{(-2.9, +2.9)}$. The $p$-values are then even worse ($0.03 \rightarrow 0.021$ and $0.03 \rightarrow 0.019$, respectively). Second, keeping the samples as many as possible can preserve the nature of human behaviors. That is why we balance the trade-off by finding the combinations of outliers as above. 

\subsection{Factor Extraction}
We now apply the multiple linear regression on $D_{C1}$ and $D_{C2}$. The affecting factors are determined based on the $p$-values with 3 significant levels: 
\begin{itemize}
\item $p \leq 0.001$: significant affecting factors
\item $0.001 < p \leq 0.01$: nearly-significant affecting factors
\item $0.01 < p \leq 0.05$: normal affecting factors
\end{itemize}

The result is described in Table~\ref{table:result}. For $D_{C1}$, we found two normal factors, including $\mathsf{thu}$ and $\mathsf{fri}$ with positive and negative coefficients, respectively. It indicates that people tend to keep their movement lifestyle on Thursday but tend to lose their movement lifestyle on Friday. For $D_{C2}$, we found one nearly-significant factor $\mathsf{fri}$ with negative coefficient like $D_{C1}$ and one normal factor $\mathsf{wk2}$ with a positive coefficient. It indicates that people tend to lose the movement lifestyle on Friday and tend to keep the movement lifestyle on the days in the second week of the month. 

\begin{table*}[!ht]
\centering
\caption{Evaluation Result}
\begin{tabular}{l r l | r r r l r r}
\textbf{case} & \textbf{no.} & \textbf{variables} & \textbf{coef} & \textbf{SE} & $\pmb{t}$ & $\pmb{p}$ & \multicolumn{2}{c}{\textbf{CI}} \\
& & && &  &  & \textbf{[0.025} &   \textbf{0.975]}\\
\hline
& 0 & (intercept) &  -0.85 & 2.53 & -0.33 & 0.74 & -5.88 & 4.19 \\
&1 & $\mathsf{mon}$  &  -20.37 & 32.36 &-0.63 &0.53 &-84.72 &43.97  \\
&2 & $\mathsf{tue}$  &  -8.38 &41.38& -0.2 &0.84& -90.65& 73.89\\
&3 & $\mathsf{wed}$ & -16.20 & 39.87& -0.41 &0.69& -95.48 &63.07  \\
&\color{red}\textbf{4}  &\color{red}\textbf{$\mathsf{thu}$}  &  \color{red}\textbf{88.72} &\color{red}\textbf{38.51} &\color{red}\textbf{2.30}& \color{red}\textbf{0.02} (*) &\color{red}\textbf{12.14} &\color{red}\textbf{165.29} \\
Exclude&\color{red}\textbf{5} & \color{red}\textbf{$\mathsf{fri}$} & \color{red}\textbf{-85.02} &\color{red}\textbf{34.67}& \color{red}\textbf{-2.45}& \color{red}\textbf{0.02} (*)& \color{red}\textbf{-153.95} & \color{red}\textbf{-16.08} \\
outlier &6 & $\mathsf{sat}$ & 40.12 &23.50& 1.71& 0.09 & -6.60& 86.84  \\
(1, 21, 94)&7 & $\mathsf{sun}$  & -24.26 &27.93& -0.87& 0.39& -79.80& 31.27\\
&8 & $\mathsf{wk1}$  & 40.35 & 54.52 &0.74& 0.46& -68.04& 148.74 \\
&9 & $\mathsf{wk2}$ & 35.21& 21.79& 1.62& 0.11& -8.12 &78.54\\
&10 & $\mathsf{wk3}$  & 4.13 &16.45& 0.25& 0.80& -28.59 &36.85  \\
&11 & $\mathsf{wk4}$  & 23.42 &23.41& 1.00& 0.32& -23.13& 69.97 \\
&12 & $\mathsf{natl}$ & -6.28& 14.36& -0.44& 0.66& -34.83& 22.27 \\
&13 & $\mathsf{wknd}$   &15.85& 17.20& 0.92& 0.36& -18.35& 50.06  \\
\hline
&0 & (concept) & -2.08 &2.62& -0.79& 0.43& -7.28& 3.13 \\
&1 & $\mathsf{mon}$  & -30.81& 32.21& -0.96& 0.34& -94.85& 33.23   \\
&2 & $\mathsf{tue}$  & -5.68 &40.62& -0.14& 0.89& -86.45& 75.08 \\
&3 & $\mathsf{wed}$ & -14.43& 38.46& -0.38& 0.71 & -90.89& 62.04  \\
&4  &$\mathsf{thu}$   &  64.42 &41.36& 1.56& 0.12& -17.81 &146.64       \\
Exclude&\color{red}\textbf{5} & \color{red}\textbf{$\mathsf{fri}$} & \color{red}\textbf{-87.45}& \color{red}\textbf{34.38} &\color{red}\textbf{-2.54}& \color{red}\textbf{0.01} (**)& \color{red}\textbf{-155.81}& \color{red}\textbf{-19.09} \\
outlier&6 & $\mathsf{sat}$  & 36.45 &23.07& 1.58& 0.12 &-9.41& 82.31  \\
(1, 21, 30) &7 & $\mathsf{sun}$  & -24.76 &27.68& -0.90& 0.37 &-79.79 &30.26\\
&8 & $\mathsf{wk1}$   & 73.12& 59.83& 1.22 &0.23& -45.84 &192.09 \\
&\color{red}\textbf{9} & \color{red}\textbf{$\mathsf{wk2}$} &\color{red}\textbf{47.83}& \color{red}\textbf{22.54}& \color{red}\textbf{2.12}& \color{red}\textbf{0.04} (*)& \color{red}\textbf{3.02} &\color{red}\textbf{92.65}\\
&10 & $\mathsf{wk3}$  & 18.33 &17.62& 1.04& 0.30& -16.70& 53.36  \\
&11 & $\mathsf{wk4}$  & 28.89& 23.57& 1.23& 0.22& -17.98& 75.76 \\
&12 & $\mathsf{natl}$& -14.40& 15.55& -0.93& 0.36& -45.32 &16.53 \\
&13 & $\mathsf{wknd}$  & 11.68& 16.80& 0.69 &0.49& -21.72 &45.09 \\
\end{tabular}\\$ $\\
Abbreviations: coef (coefficient), SE (standard error), CI (confidence interval)
\label{table:result}
\end{table*}

\section{Discussion}
\label{section:discussion}
The result can be heuristically explained that Thursday and the second week of the month are the middle time of the week and the month, respectively. The human behavior (even human mood and social interaction) becomes more stable than the first days of the weeks (the first weeks of the month) and the days near to the weekends (weeks near to the end of the month). In contrast, the result of Friday may be caused by ``nomikai'' which is a drinking party (often on Friday and with co-workers) phenomenon particular to Japanese culture. Even so, a deeper analysis and formal proof of these results should be investigated for future work. 

Our findings can help understand more about human mobility psychological and behavioural science which is important for urban planning, traffic forecasting, and the spread of biological and mobile viruses. They can also help enhance the effectiveness of the location-based recommendations and the location-based predication, and enable the advertisers to design and present their location services to targeted customers. For example, if a restaurant where a customer visited before knows that he tends to lose the habit of going to the restaurant on Friday, it can promote more discounts on those days rather than the other days of the week. 

In this paper, weekly temporal patterns were analyzed. Future work can examine daily temporal patterns which are different time frames during a day such as $\{$00:01-06:00, 06:01-12:00, 12:01-18:00, 18:01-24:00$\}$  (for the interval of 6 hours),  $\{$00:01-03:00, 03:01-06:00, $\cdots$, 21:01-24:00$\}$ (for the interval of 3 hours), etc. Combining weekly and daily temporal patterns is also a promising approach to figure out which time bins that people tend to visit (or tend to lose the habit of visiting) the usual location.

\section{Conclusion}
\label{section:conclusion}
In this paper, we aim to find which temporal factors that affect the human mobility lifestyle. We collected GPS data including longitude, latitude, and timestamp from 100 random participants in Japan using a smartphone application. We designed a regression model that utilizes 13 weekly temporal factors as independent variables categorized into 3 pattern types: days of the week, quarters of the month, and holidays. We proposed an algorithm to compute the similarity score between the location history and the most recent location log. We applied a multiple linear regression with a $t$-test and found that people tend to keep their mobility habit on Thursday and the days in the second week of the month but tend to lose the habit on Friday.

\end{document}